\documentclass[aps,pra,twocolumn,showpacs,superscriptaddress]{revtex4}
\usepackage{bm,amsmath,amssymb,latexsym,mathrsfs,graphicx,enumerate}
\usepackage[mathcal]{euscript}
\usepackage{hyperref}
\usepackage{epsfig}

\usepackage{amsmath}
\usepackage{graphics}
\begin{document}

\title{\textbf{On Stability of Flat Band Modes in a Rhombic Nonlinear Optical Waveguide Array}}

\author{Andrey I. Maimistov}
\affiliation{\normalsize \noindent Department of Solid State Physics
and Nanostructures, National Nuclear Research University,
Moscow Engineering Physics Institute, Moscow, 115409,}
\affiliation{\normalsize \noindent Department of General Physics,
Moscow Institute of Physics
and Technology, Dolgoprudny, Moscow region, 141700 Russia, \\
E-mails: aimaimistov@gmail.com \\
}

\date{\today}

\begin{abstract}
\noindent  The quasi-one-dimensional rhombic array of the waveguides
is considered. In the nonlinear case the system of equations
describing coupled waves in the waveguides has the solutions that
represent the superposition of the flat band modes. The property of
stability of these solutions is considered. It was found that the
flat band solution is unstable until the power threshold be
attained.
\end{abstract}



\maketitle

%


\section{Introduction}

\noindent The optical simulations of the different phenomena of the
condensed mater physics
\cite{Yannopapas:14,Onoda:09,Dragom:14,He:14}, the quantum physics
\cite{Longi:09,Rodri:14}, and the cosmology
\cite{Greenl:07,Nariman:09,Smolyan:12,Smolyan:13} are the object of
resent investigation. Under consideration of the 2D electron systems
it was found that the presence of the third atom in the unit cell of
the lattice leads to emerging of a flat band between conventional
energy zones. Similar optical lattices can be realized by means of
waveguides as nodes of the lattice
\cite{Vicencio:14,Fang:15,Vicencio:Cantillano:15}. Some kinds of the
optical lattices that demonstrate the photonic spectrum with flat
band have been discussed in \cite{Longi:14,Maimis:15,Maim:Gabi:16}.
If the electric field in the optical lattice is made up from the
flat band modes then this field demonstrates the diffractionless
propagation along waveguide array.

\begin{figure}[h]
    \centering
   \includegraphics[scale=0.35]{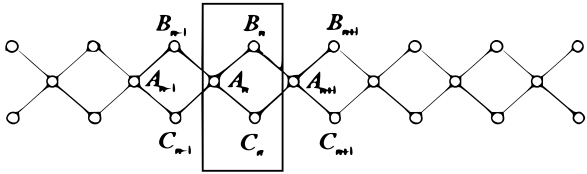}
    \caption{The rhombic array of waveguides. The unit cell is shown by
    rectangular box. }
    \label{Romb:array:1}
\end{figure}

Recently the quasi-one-dimensional array of the waveguides
containing the three linear chain of waveguides was considered
\cite{Longi:14}. The central chain marked as A-type chain is placed
between two chains of waveguides, which are marked as B- and C- type
ones. These chains of waveguides are shifted relative to the A-chain
at one half of period of the lattice. Interaction between waveguides
is due to tunnel coupling. Furthermore, coupling between A-B and A-C
waveguides takes place only. This waveguide system alike to a double
zigzag array or to an array of rhombus (see Fig.
\ref{Romb:array:1}).

This waveguide array named as the quasi-one-dimensional rhombic
array has been studied in
\cite{Longi:14,Mukherjee:15,Mukherjee:Spracklen:15a} in the case of
linear optical waveguides.

The purpose of this paper is to study the stability of the
electromagnetic field distribution in the quasi-one-dimensional
rhombic array of the nonlinear waveguides. The nonlinearity is
described by the susceptibility of third order. In Sec.
 \ref{sec:solutions} the some nonlinear analogies of the superposition of the
flat band modes are found. The discrete diffraction for these
electric field distributions over the waveguides is absent.  The
stability of these solutions is investigated by the use of the
linear stability analysis in Sec. \ref{sec:stabil}. Taking into
account that the field distribution over the waveguide is the
superposition of the band modes, we can make inferences about
stability band mode from analysis of the stability properties of the
field distribution.

\section{Model and basic equations}\label{sec:equations}

\noindent It is assumed that all waveguides are manufactured from a
nonlinear dielectric, which is characterized by Kerr nonlinearity.
System of equations describing coupled waves in this
quasi-one-dimensional rhombic nonlinear optics waveguide array
(RNOWA) has the following form
\begin{eqnarray}
&& i\left(\frac{\partial}{\partial\tau}+ \frac{\partial}{\partial
\zeta} \right)A_n =
c_1(B_{n}+B_{n-1})+ \nonumber\\
&& \qquad \qquad \qquad \qquad +c_2(C_{n}+C_{n-1}) + \mu_1|A_n|^2A_n, \nonumber\\
&& i\left(\frac{\partial}{\partial \tau}+ \frac{\partial}{\partial
\zeta} \right)B_{n} = c_1(A_{n+1}+A_{n})
+ \label{eq:romb:nlin:ABCn:1} \\
&& \qquad \qquad \qquad \qquad + \mu_2|B_n|^2B_n, \nonumber\\
&& i\left(\frac{\partial}{\partial \tau}+ \frac{\partial}{\partial
\zeta} \right)C_{n} = c_2(A_{n+1}+A_{n}) +\mu_3|C_n|^2C_n.
 \nonumber
\end{eqnarray}
Here $A_n $, $B_n $ and $C_n $ are dimensionless slowly varying
amplitudes of the electric fields propagating in waveguides of
RNOWA. The sub-indices are numbering unit cells (see
Fig.\ref{Romb:array:1}). It is assumed that the phase matching
condition is satisfied. The coefficients $c_1$ and $c_2$ specify the
coupling between waveguides from different chains. The parameters
$\mu_1$, $\mu_2$ and $\mu_3$ represent the self-interaction effect
in waveguides. If $c_1=c_2$ and $\mu_1 =\mu_2=\mu_3 =0$, the system
of equations (\ref{eq:romb:nlin:ABCn:1}) is reduced to the system of
the linear equations considered in
\cite{Mukherjee:15,Mukherjee:Spracklen:15a}. The symmetric rhombic
array, where $c_1=c_2=1$ and $\mu_b=\mu_c $ will be considered. This
contraction allows us to reduce the system of equations
(\ref{eq:romb:nlin:ABCn:1}) to following equations
\begin{eqnarray}
&& i\frac{\partial}{\partial\zeta} A_n =
(B_{n}+B_{n-1})+(C_{n}+C_{n-1}) +\mu_1|A_n|^2A_n, \nonumber\\
&& i\frac{\partial}{\partial \zeta}B_n  = (A_{n+1}+A_{n}) +
 \mu_2|B_n|^2B_n, \label{eq:romb:nlin:ABCn:sim:2} \\
&& i\frac{\partial}{\partial \zeta} C_{n} = (A_{n+1}+A_{n})+
 \mu_2|C_n|^2C_n. \nonumber
\end{eqnarray}

If the power per one unit cell is defined by expression
$W_n=|A_n|^2+|B_n|^2+|C_n|^2$, the equation
$$
\frac{\partial W_n}{\partial\zeta}+i\left[A_{n+1}(B_n^{*}+B_n^{*})+
A_n^{*}(B_{n-1}+C_{n-1})- c.c. \right] =0.
$$
can be obtained from the system of equations
(\ref{eq:romb:nlin:ABCn:sim:2}). The local value
\begin{equation}\label{eq:romb:nlin:curent:1}
   f_n= i(A_nD^{*}_{n-1}-A_n^{*}D_{n-1}),
\end{equation}
is introduced, where $D_n=B_n+C_n$. So, the equation for $W_n$ can
be rewritten now as
\begin{equation}\label{eq:romb:nlin:conserv:1}
   \frac{\partial W_n}{\partial \zeta} +(f_{n+1}-f_n) =0.
\end{equation}
The expression in brackets can be interpreted as the discrete
divergence of the current density $f_n$ in 1D space. The equation
(\ref{eq:romb:nlin:conserv:1}) is the discrete continuity equation.

There are three constrains: (a) $A_n=0$, $B_n=-C_n$, (b)
$A_n=(-1)^nA$, $B_n=(-1)^nB$, $C_n=(-1)^nC$, and (c) $A_n=(-1)^nA$,
$B_n=(-1)^nB$, $C_n=(-1)^{n+1}C$. For all these constrains the
(\ref{eq:romb:nlin:conserv:1}) is hold. However, in the case of (a)
and (c) the current density $f_n$ is zero for any $A$, $B$ and $C$.
In the case of (b) $f_n=(-i)(AD^{*}-A^{*}D)$. This current will be
zero only if $A=0$. In the linear case the distribution of the kind
(a) corresponds to the flat band modes of
\cite{Longi:14,Mukherjee:15,Mukherjee:Spracklen:15a}. In the RNOWA
the electric field distributions under the constrains (a) and (c)
can be considered as a nonlinear version of the superposition of the
flat band modes. In following the electric field distribution (a)
will be studied.

\section{The flat band solutions}\label{sec:solutions}

\noindent With taking into account constrain $A_n=0$, $B_n=-C_n$,
the system of equations (\ref{eq:romb:nlin:ABCn:sim:2}) can be
represented as
\begin{eqnarray}
  && i\frac{\partial}{\partial \zeta} A_n =0, \quad
  i\frac{\partial}{\partial \zeta} B_n =  \mu_2|B_n|^2B_n,\nonumber  \\
  && i\frac{\partial}{\partial \zeta} C_n =  \mu_2|C_n|^2C_n. \label{eq:romb:nlin:ABCn:sim:An0:1}
\end{eqnarray}

Defining the real variables $a_n$, $b_n$, $c_n$, $\varphi_a$,
$\varphi_b$ and $\varphi_c$ from the formulas
$A_n=a_n\exp(i\varphi_a)$, $B_n=b_n\exp(i\varphi_b)$ and
$C_n=c_n\exp(i\varphi_a)$, the real equations can be derived
$$
\frac{\partial a_n}{\partial \zeta} = 0,\quad \frac{\partial
b_n}{\partial \zeta} = 0,\quad \frac{\partial c_n}{\partial \zeta} =
0,
$$
$$
\frac{\partial \varphi_{a n}}{\partial \zeta} = -\mu_1a_n^2, \quad
\frac{\partial \varphi_{b n}}{\partial \zeta} = -\mu_2b_n^2, \quad
\frac{\partial \varphi_{c n}}{\partial \zeta} = -\mu_2c_n^2.
$$
As the case of $A_n=0$ is considered the phase $\varphi_{a n}$ is
indeterminate. The amplitudes $b_n$ and $c_n$ are constant, $b_{n0}$
and $c_{n0}=-b_{n0}$.

With taking into account this result the solutions of these
equations can be written as
$$
\varphi_b = \varphi_c = -\mu_2 b_{n0}^2\zeta.
$$
The initial phases are constants of integration and can be chosen to
be zero. Thus the solution of the the system of equations
(\ref{eq:romb:nlin:ABCn:sim:An0:1}) reads as
\begin{equation}\label{eq:romb:nlin:flatband:sim:1}
    \tilde{A}_n =0, \quad \tilde{B}_n = b_{n0}e^{-i\mu_2 b_{n0}^2\zeta}, \quad \tilde{C}_n
    = -b_{n0}e^{-i\mu_2 b_{n0}^2\zeta}.
\end{equation}

The electric field distribution (\ref{eq:romb:nlin:flatband:sim:1})
characterized by the diffractionless propagation along waveguides
will referred to as flat band solution of the equation
(\ref{eq:romb:nlin:ABCn:sim:2}).

\section{Stability of the flat band solution}\label{sec:stabil}

\noindent Let us consider the homogeneous distribution of the
electric field amplitudes over waveguide array $b_{n0}= b_{0}$. The
stability of the solution (\ref{eq:romb:nlin:flatband:sim:1}) can be
analyzed by introducing small perturbations into the electric field
amplitudes:
\begin{eqnarray}
  && A_n=  p_ne^{-i\mu_2b^2_{0}\zeta}, \nonumber\\
 && B_n=\tilde{B}_n  + b_n = (b_{0}+q_n)e^{-i\mu_2b^2_{0}\zeta},  \nonumber\\
  && C_n=\tilde{C}_n +c_n= (-b_{0}+r_n)e^{-i\mu_2b^2_{0}\zeta},   \nonumber
\end{eqnarray}
where $p_n$, $q_n$ and $r_n$ are the small perturbations of the
fields in $n$-th unit cell.

The linearized system of equation for these perturbations takes the
form
\begin{eqnarray}
&& i\frac{\partial p_n}{\partial \zeta} = -\varrho p_n+(q_{n}+q_{n-1})+(r_{n}+r_{n-1}), \nonumber\\
&& i\frac{\partial q_n}{\partial \zeta}  = (p_{n}+p_{n+1}) +
 \varrho(q_n + q_n^{*}), \label{eq:romb:nlin:flatband:sim:4} \\
&& i\frac{\partial r_n}{\partial \zeta} = (p_{n}+p_{n+1})+
 \varrho(r_n + r_n^{*}). \nonumber
\end{eqnarray}
Here $\varrho =\mu_2b^2_{0}$.

Let there be $N=2M+1$ waveguides in RNOWA. The fields in $n$-th unit
cell are presented as the Fourier serieses
\begin{eqnarray}
  && p_n= \sum_{s=-M}^{s=M}\left(p_se^{2\pi i ns/M}+\bar{p}_se^{-2\pi i ns/M}\right), \nonumber \\
  && q_n= \sum_{s=-M}^{s=M}\left(q_se^{2\pi i ns/M}+\bar{q}_se^{-2\pi i ns/M}\right),
  \label{eq:romb:nlin:discrFour} \\
  && r_n= \sum_{s=-M}^{s=M}\left(r_se^{2\pi i ns/M}+\bar{r}_se^{-2\pi i ns/M}\right). \nonumber
\end{eqnarray}
Substitution of the (\ref{eq:romb:nlin:discrFour}) in the equations
(\ref{eq:romb:nlin:flatband:sim:4}) with taking into account the
orthogonality of the harmonic functions 
results in the following equations
for modes
\begin{eqnarray}
&& i\frac{\partial p_s}{\partial \zeta} = -\varrho
p_s+\kappa(s)^{*}(q_s +r_s), \nonumber\\
&& i\frac{\partial \bar{p}_s}{\partial \zeta} = -\varrho \bar{p}_s+\kappa(s)(\bar{q}_s +\bar{r}_s), \nonumber\\
&& i\frac{\partial q_s}{\partial \zeta}  = \kappa(s)p_{s} +
 \varrho(q_s + \bar{q}_s^{*}), \nonumber\\
&& i\frac{\partial \bar{q}_s}{\partial \zeta}  =
\kappa(s)^{*}\bar{p}_{s} + \varrho(\bar{q}_s + q_s^{*}), \nonumber\\
&& i\frac{\partial r_s}{\partial \zeta} = \kappa(s)p_{s} +
 \varrho(r_s + \bar{r}_s^{*}), \nonumber\\
&& i\frac{\partial \bar{r}_s}{\partial \zeta} =
\kappa(s)^{*}\bar{p}_{s} +  \varrho(\bar{r}_s + r_s^{*}).\nonumber
\end{eqnarray}
Here
$$
\kappa(s) = 2\cos(\pi s /M)e^{i\pi s/M}.
$$
In following the mode marks $s$ can be omitted, as long as it will
be necessary to indicate mode marks. If the new functions $w =
\kappa p, ~ \bar{w} = \kappa^{*} \bar{p} $ are introduced, this
system of equations will take the form
\begin{eqnarray}
&& i\frac{\partial w}{\partial \zeta} = -\varrho w+|\kappa|^{2}(q
+r), \quad i\frac{\partial \bar{w}}{\partial \zeta} = -\varrho \bar{w}+|\kappa|^2(\bar{q} +\bar{r}), \nonumber\\
&& i\frac{\partial q}{\partial \zeta} = w + \varrho(q +
\bar{q}^{*}),~~~\quad \quad i\frac{\partial \bar{q}}{\partial \zeta}
=
\bar{w} + \varrho(\bar{q} + q^{*}), \label{eq:romb:nlin:flatband:sim:5} \\
&& i\frac{\partial r}{\partial \zeta} = w + \varrho(r +
\bar{r}^{*}), ~~~\quad \quad i\frac{\partial \bar{r}}{\partial
\zeta} = \bar{w} + \varrho (\bar{r} + r^{*}). \nonumber
\end{eqnarray}
From these equations one can find the closed system of three
equations
\begin{eqnarray}
  && -\frac{\partial^2 q}{\partial \zeta^2} = -\varrho \bar{w}^*+ |\kappa|^{2}(q
+r), \nonumber \\
  && -\frac{\partial^2 r}{\partial \zeta^2} = -\varrho \bar{w}^*+ |\kappa|^{2}(q
+r),\nonumber \\
&&  -\frac{\partial^2 \bar{w}^*}{\partial \zeta^2}
=(2|\kappa|^{2}+\varrho^2) \bar{w}^*+\varrho |\kappa|^{2}(q +r).
\nonumber
\end{eqnarray}
If the variable $u=q+r$, $\tilde{u}=q-r$, $\bar{w}^* =v$ are used,
the either system of equation can be written
\begin{eqnarray}
  &&\frac{\partial^2 v}{\partial \zeta^2} + (2|\kappa|^{2}+\varrho^2)v +\varrho |\kappa|^{2}u=0, \nonumber \\
  && \frac{\partial^2 u}{\partial \zeta^2} - 2\varrho v +2 |\kappa|^{2}u =0, \label{eq:romb:nlin:flatband:sim:7} \\
  && \frac{\partial^2 \tilde{u}}{\partial \zeta^2}=0. \nonumber
\end{eqnarray}
Thus the variable $\tilde{u}=q-r$ varies as $\tilde{u}=
\tilde{u}_0+\tilde{u}_1 \zeta$. Hence, the small perturbations vary
proportionally with the distance $\zeta$. In this since the flat
band solution is unstable. However, it is a weak instability.

If the initial variable $q$, $r$ è $v =\bar{w}^*$ are considered,
the corresponding characteristic equation takes the form
$$
\left|
\begin{array}{ccc}
  \lambda^2+(2|\kappa|^{2}+\varrho^2)  & \varrho |\kappa|^{2} & \varrho |\kappa|^{2} \\
  -\varrho & \lambda^2+2|\kappa|^{2} & |\kappa|^{2} \\
  -\varrho & |\kappa|^{2} & \lambda^2+2|\kappa|^{2} \\
\end{array}\right| =0.
$$
The calculation of this determinant results in the algebraic
equation
\begin{equation}\label{eq:romb:nlin:flatband:sim:determ}
    \lambda^2\left[(\lambda^2+ 2|\kappa|^{2}+\varrho^2)(\lambda^2+2|\kappa|^{2}) +
     2\varrho^2|\kappa|^{2} \right]=0.
\end{equation}

The roots of equation (\ref{eq:romb:nlin:flatband:sim:determ})
$\lambda^2=0$ are evidence for the linear increasing o the small
perturbations. Another roots can be found from reduced
characteristic equation
\begin{equation}\label{eq:rhomb:chara:1}
    (\lambda^2+ 2|\kappa|^{2}+\varrho^2)(\lambda^2+2|\kappa|^{2}) +
     2\varrho^2|\kappa|^{2} =0.
\end{equation}
The changing $\lambda^2=2|\kappa|^{2}\xi$ results in equation $
(1+\xi)(1+\xi+\mu)+\mu =0$, where $\mu=\varrho^2/(2|\kappa|^2)$. It
follows that roots of this equation read as
$$
\xi_{1,2}= -\left(1+\frac{\mu}{2} \right) \pm \sqrt{D_e},
$$
where $D_e=\mu^2/4 -\mu$. Thus, the roots of equation
(\ref{eq:rhomb:chara:1}) are given by the expressions
\begin{equation}\label{eq:nlin:flatband:sim:roots}
    \lambda_{1,2}^{(\pm)}=\pm \sqrt{2|\kappa|}\left[-\left(1+\frac{\mu}{2} \right) \pm \sqrt{D_e}
    \right]^{1/2}.
\end{equation}
The instability takes place if $\mathrm{Re}(\lambda_{1}^{(\pm)})>0$
or/and $\mathrm{Re}(\lambda_{2}^{(\pm)})>0$.

In a linear case, where $\mu=0$, $\lambda_{1,2}^{(\pm)}=\pm
2i|\kappa|$. It means that there is no the exponential increasing of
the perturbations. However, any small perturbations lead to
spreading the electromagnetic waves in transversal direction. It is
due to $\lambda^2=0$. Thus the discrete diffraction in a linear 1D
rhombic waveguide array takes place \cite{Maim:Patrik:16}.

The roots of the equations (\ref{eq:rhomb:chara:1}) can be written
as
$$
\lambda_{1}^{(\pm}=\pm\sqrt{2}|\kappa|\sqrt{\xi_1}, ~~
\lambda_{2}^{(\pm)}=\pm \sqrt{2}|\kappa|\sqrt{\xi_2}.
$$

If $0<\mu<4$ the discriminant $D_e$ is negative one,
$D_e=-\mu(4-\mu)/4$, hence
$$
\xi_{1,2}= -\left(1+\frac{\mu}{2} \right) \pm i\sqrt{|D_e|}.
$$

By extracting square root from $\xi_{1,2}$ one can obtain the
expression for roots of the equation (\ref{eq:rhomb:chara:1}):
\begin{eqnarray}
  && \lambda^{(\pm)}_1=\Lambda (\cosh\phi^{+}+i\sinh\phi^{+}), \nonumber  \\
  && \lambda^{(\pm)}_2=\Lambda (\cosh\phi^{-}+i\sinh\phi^{-}), \nonumber
\end{eqnarray}
where
$$\Lambda = \sqrt{2} |\kappa| \left(1+\frac{\mu}{2} \right)^{1/2}, \quad
  \sinh2\phi=\pm \frac{\sqrt{|D_e|}}{1+\mu/2}.
$$

As $\mathrm{Re}(\lambda_{1}^{(\pm)})> 0$ and
$\mathrm{Re}(\lambda_{2}^{(\pm)})< 0$, the flat band solution under
consideration is unstable in the region $0<\mu<4$.

If $\mu > 4$ the discriminant $D_e=\mu(\mu-4)/4$ is positive. Hence
$\xi_{1,2}$ is real value and
$$ \xi_{1}= -\left(1+\frac{\mu}{2}
\right) + \sqrt{D_e},\quad \xi_{2}= -\left(1+\frac{\mu}{2} \right) -
\sqrt{D_e}.
$$
For $\xi_{1}$ the following expression
$$
\xi_{1}= -\left(1+\frac{\mu}{2} \right)
+\frac{\mu}{2}\sqrt{1-\frac{4}{\mu^2}}= -1-\frac{\mu}{2}\left(1-
\sqrt{1-\frac{4}{\mu^2}} \right) ,
$$
can be found. It is negative at $\mu > 4$. From the definition
$\xi_{2}$ it follows that $\xi_{2}<0$. Thus,
$\mathrm{Re}(\lambda_{1}^{(\pm)})= 0$ è
$\mathrm{Re}(\lambda_{2}^{(\pm)})= 0$. Hence, the flat band solution
is stable in the region $\mu >\mu_c =4$.

So, the flat band solution (\ref{eq:romb:nlin:flatband:sim:1}) is
unstable if the radiation power in waveguide is less than some
threshold power. This solution will be stable if the power is more
than threshold value.

Using the definition of the relevant parameters
$$
\varrho =\mu_2b^2_{0}, \quad |\kappa(s)| = 2|\cos(\pi s /M)|,
$$
one can write the stability condition in the form
\begin{equation}\label{eq:nlin:flatband:sim:instab:1}
    \mu_2b_0^2 \geq 2\sqrt{2}|\cos(\pi s /M)|.
\end{equation}
If the normalized power per mode $b_{0}$ will be greater than the
critical value ($b_{0c}^2 = 2\sqrt{2}\mu_2^{-1}|\cos(\pi s /M)|$)
the perturbations will not increase exponentially. It should be
mentioned that the critical value $b_{0c}$ is depended on mode
marker $s$. Hence only part of the modes having markers, which are
belong to the interval
$$
\frac{\pi}{2}>\frac{\pi s}{M} \geq \arccos
\frac{\mu_2b_0^2}{2\sqrt{2}},
$$
will be stable. All modes of the flat band will be stable if the
condition $\mu_2b_0^2\geq 1$ is held.

 \section{Conclusion}

\noindent The rhombic nonlinear optical waveguide array is
considered in the paper. In the case of linear waveguides the
waveguide array of this kind has been investigated in
\cite{Longi:14,Mukherjee:15,Mukherjee:Spracklen:15a}. It was shown
that all (normal) mode of this waveguide array are separated on
three groups or bands in 1D space of the wave vectors. The two bands
are populated by the modes describing the discrete diffraction in
waveguide array.  Third band contains the modes that describes the
wave propagating without diffraction. This  band was named as the
flat band.

In the RNOWA the flat band analog exists. There is the solutions of
the system of equation of the RNOWA describing the diffractionless
waves propagation. However, both the linear and nonlinear the flat
band solution are weak instable. The small perturbations increase
directly with the first power of distance along the waveguide. In
the case of nonlinear waveguides the small perturbations grow
exponentially. But with power increasing the part of modes began be
stable. The all modes became stable if the power of radiation in
waveguide is greater then the threshold value. Theses flat band
solution are stable but not asymptotic stable. The phenomenon is
similar to the self-focusing in a nonlinear bulk medium.

As it was pointed above, there are two kind of the electric field
distributions: (a) $A_n=0$, $B_n=-C_n$ and (b) $A_n=(-1)^nA$,
$B_n=(-1)^nB$, $C_n=(-1)^{n+1}C$, for which the power flux between
unit cells of the RNOWA is equal to zero. Here the case of (a) was
investigated. As for the second case, there is discrete analogy of
the modulation instability for a cubic nonlinear bulk material.
There is threshold power for each mode and the threshold power for
all band. If the  power of radiation in waveguide of RNOWA is
greater than this threshold the perturbations are increased
exponentially with distance.

\section*{ Acknowledgement}

I am grateful to Prof. I. Gabitov and Dr. C. Bayun for enlightening
discussions. This investigation is funded by Russian Science
Foundation (project 14-22-00098).

\end{document}